\documentclass[aps,onecolumn,showpacs,footinbib]{revtex4}%
\usepackage[latin9]{inputenc}
\usepackage{amssymb}
\usepackage{amsmath}
\usepackage{amscd}
\usepackage{graphicx}
\usepackage{subfigure}
\usepackage{float}
\begin{document}
\draft
\title{Arbitrary control of multiple-qubit systems in the symmetric Dicke subspace}
\author{Shi-Biao Zheng\thanks{%
E-mail: sbzheng@pub5.fz.fj.cn}}
\address{Department of Electronic Science and Applied Physics\\
Fuzhou University\\
Fuzhou 350002, P. R. China}
\date{\today }

\begin{abstract}
We discuss a general physical mechanism for arbitrary control of the quantum
states of multiple qubits in the symmetric Dicke subspace. The qubit-qubit
coupling leads to unequal energy spacing in the symmetric Dicke subspace.
This allows one to manipulate a prechosen transition with an external
driving source, with other transitions remaining off-resonant. Any entangled
state in the symmetric Dicke subspace can be created from the initial ground
state by tuning the driving source. We illustrate the idea in cavity QED,
but it should be applicable to other related systems.
\end{abstract}

\pacs{PACS number: 03.67.Mn, 03.65.Ud, 42.50.Dv}

\vskip 0.5cm \maketitle

\narrowtext

\section{INTRODUCTION}

The control of quantum systems is fundamental to quantum optics as any
experimental investigation of nonclassical features relies on the ability to
create and manipulate quantum states. Mathematically, one can design an
unitary evolution operator to transform an initial state to the desired
state. In physics, the main difficulty comes from the fact that the desired
evolution operator is limited by the attainable Hamiltonian. In the context
of cavity QED, methods has been presented to force an quantized
electromagnetic field localized in a cavity from an initial ground state to
any quantum state [1-3]. In these schemes the quantum state of one or more
atoms are manipulated in an controllable way and the coherence of the atom
is transferred to the cavity field. The atoms act as the source, $^{^{\prime
\prime }}$teaching$^{^{\prime \prime }}$ the cavity field to evolve to the
desired state.

For a system composed of multiple particles, the superpositions of product
state leads to entanglement. There are various types of multi-particle
entanglement and the characterization of entanglement has not been completed
yet. The control of quantum states of composite systems is a prerequisite
for experimental study of entanglement properties [4,5]. Besides fundamental
interest, the control of the time evolution of multi-particle systems is of
importance for the implementation of quantum computers. The implementation
of a quantum computational task corresponds to the performance of an unitary
transformation on the quantum register, which is composed of multiple
quantum-bits (qubits) [6]. During the quantum logic operation, the qubits
are generally in an entangled state. In essence, realizing a quantum
computer is equivalent to controlling the time evolution of an N-qubit
system. The Hilbert space increases exponentially as the number of qubits
increases and the control of multi-qubit systems is very complex. As far as
we know, no realistic mechanism has been proposed for creating an arbitrary
entangled state for N-qubit systems.

In this paper we design a general interaction Hamiltonian which can drive an
N-qubit system to evolve from an initial product state to any superposition
state in the symmetric Dicke subspace. The qubit-qubit coupling leads to
unequal spacing of energy levels in the symmetric Dicke subspace. This
allows one to arbitrarily manipulate a specific transition by using an
external driving field, which is resonant with the prechosen transition but
off-resonant with other transitions. Any symmetric entangled state can be
obtained by appropriately adjusting the parameters of the external driving
field. The well-known Greenberger-Horne Zeilinger (GHZ) states [7] and W
states [8] are special classes of symmetric states. Our idea provides a
possibility for the creation of general symmetric entangled states. The idea
can be realized in realistic physical systems.

The paper is organized as follows. In Sec.2, we describe the interaction
Hamiltonian for the manipulation of a prechosen transition in the symmetric
Dicke subspace. In Sec.3, we show that any symmetric entangled state can be
created from the initial ground state by tuning the parameter of the
external souce. In Sec.4, we discuss the physical realization of the scheme
in cavity QED and analyse decoherence effects. We also estimate the
probability that the atomic system undergoes unwanted transitions. A summary
appears in Sec.5.

\section{THE INTERACTION HAMILTONIAN}

We consider an N-qubit system. The qubits have two states $\left|
e\right\rangle $ and $\left| g\right\rangle $ with energy-level difference $%
\omega _0$. The Hamiltonian for the whole system is (assuming $\hbar =1$)
\begin{equation}
H=H_0+H_1+H_2,
\end{equation}
where
\begin{equation}
H_0=\omega _0S_z,
\end{equation}
\begin{equation}
H_1=\lambda S^{+}S^{-},
\end{equation}
\begin{equation}
H_2=\varepsilon [e^{-i(\omega t-\theta )}S^{+}+e^{-i(\omega t-\theta
)}S^{-}],
\end{equation}
\begin{eqnarray}
S_z &=&\frac 12\sum_{j=1}^N(\left| g_j\right\rangle \left\langle g_j\right|
-\left| e_j\right\rangle \left\langle e_j\right| ), \\
S^{+} &=&\sum_{j=1}^N\left| e_j\right\rangle \left\langle g_j\right| ,
\nonumber \\
S^{-} &=&\sum_{j=1}^N\left| g_j\right\rangle \left\langle e_j\right| .
\nonumber
\end{eqnarray}
$H_1$ describes the qubit-qubit coupling with coupling strength $\lambda $,
and $H_2$ describes the interaction between the qubit system and an external
driving source with frequency $\omega $. The parameters $\omega ,$ $%
\varepsilon $, and $\theta $ are controllable via the external driving
source. The operators $S^{+}$, $S^{-}$, and $S_z$ obey the commutation
relations of the angular-momentum operators. If the system is initially in a
proper Dicke state $\left| J,-J+k\right\rangle $ [9], it would evolve within
a subspace of the Dicke space spanned by $\left\{ \left| J,-J\right\rangle
,\left| J,-J+1\right\rangle ,...,\left| J,J\right\rangle \right\} $. We
consider the case that the system is initially symmetric, thus the Hilbert
space reduces to the symmetric Dicke subspace with $J=N/2$, formed by the
Dicke states which are symmetric under the permutation of any two qubits.
The state $\left| J,-J+k\right\rangle $ is a symmetric state with $k$
particles being in the state $\left| e\right\rangle $, i.e.,
\begin{equation}
\left| J,-J+k\right\rangle =%
{2J \choose k}
^{-1/2}\sum_jP_j(\left| e_1,e_2,...,e_k,g_{k+1},g_{k+2},...g_N\right\rangle
),
\end{equation}
where \{$P_j$\} denotes the set of all distinct permutations of the qubits.
In the symmetric Dicke subspace the collective operators $S^{+}$ and $S^{-}$
act as

\begin{eqnarray}
S^{+}\left| J,-J+k\right\rangle &=&\sqrt{(2J-k)(k+1)}\left|
-J,-J+k+1\right\rangle , \\
S^{-}\left| J,-J+k\right\rangle &=&\sqrt{k(2J-k+1)}\left|
J,-J+k-1\right\rangle  \nonumber
\end{eqnarray}
The Hamiltonian $H_1$ does not induce transition between different symmetric
Dicke states, but shift the energy-level of the state $\left|
J,-J+k\right\rangle $ by $k(2J-k+1)\lambda $. Due to the qubit-qubit
coupling the energy-level spacing between $\left| N/2,M+1\right\rangle $ and
$\left| N/2,M\right\rangle $ is $\omega _0+2(J-k)\lambda $, which is
depending upon the excitation number of the state $\left|
J,-J+k\right\rangle $. Thus the spacing of energy levels in the symmetric
Dicke subspace becomes unequal. The detuning between the transition $\left|
J,-J+k\right\rangle \rightarrow \left| J,-J+k+1\right\rangle $ and the
classical source is $\omega _0+2(J-k)\lambda -\omega $.

Suppose that $\varepsilon \ll \lambda $ and $\omega =\omega _0+2(J-k)\lambda
$. In this case only the transition $\left| J,-J+k\right\rangle \rightarrow
\left| J,-J+k+1\right\rangle $ is resonant with the external driving source,
while other transitions remains far off-resonance and can be neglected.
Therefore, the symmetric Dicke subspace further reduces to $\{\left|
J,-J+k+1\right\rangle ,\left| J,-J+k\right\rangle \}$ and the Hamiltonians $%
H_0$, $H_1$, and $H_2$ reduce to
\begin{equation}
H_0=2(k-J)\omega _0\left| J,-J+k\right\rangle \left\langle J,-J+k\right|
+(2J-2k+1)\omega _0\left| J,-J+k+1\right\rangle \left\langle J,-J+k+1\right|
,
\end{equation}
\begin{equation}
H_1=\alpha _k\left| J,-J+k\right\rangle \left\langle J,-J+k\right| +\alpha
_{k+1}\left| J,-J+k+1\right\rangle \left\langle J,-J+k+1\right| ,
\end{equation}
\begin{equation}
H_2=\eta _k(e^{-i(\omega t-\theta )}\left| J,-J+k+1\right\rangle
\left\langle J,-J+k\right| +e^{i(\omega t-\theta )}\left|
J,-J+k\right\rangle \left\langle J,-J+k+1\right| ),
\end{equation}
where
\begin{eqnarray}
\alpha _k &=&k(2J-k+1)\lambda , \\
\eta _k &=&\sqrt{(2J-k)(k+1)}\varepsilon .  \nonumber
\end{eqnarray}
In the interaction picture with respect with $H_0$ we obtain the interaction
Hamiltonian
\begin{equation}
H_I=H_1+H_{2,I},
\end{equation}
where
\begin{equation}
H_{2,I}=\eta _k[e^{i2(k-J)\lambda t}\left| J,-J+k+1\right\rangle
\left\langle J,-J+k\right| +e^{-i2(k-J)\lambda t}\left| J,-J+k\right\rangle
\left\langle J,-J+k+1\right| ].
\end{equation}
Taking advantage of the unequal energy spacing induced by the qubit-qubit
coupling, we can selectively manipulate any transition in the symmetric
Dicke subspace by tuning the frequency of the external field appropritely.

\section{GENERATION OF ANY SYMMETRIC ENTANGLED STATE}

The time evolution of this system is decided by Schr\"odinger's equation:
\begin{equation}
i\frac{d|\psi (t)\rangle }{dt}=H_I|\psi (t)\rangle .
\end{equation}
Perform the unitary transformation
\begin{equation}
|\psi (t)\rangle =e^{-iH_1t}|\psi ^{^{\prime }}(t)\rangle .
\end{equation}
Then we obtain
\begin{equation}
i\frac{d|\psi ^{^{\prime }}(t)\rangle }{dt}=H_{2,I}^{^{\prime }}|\psi
^{^{\prime }}(t)\rangle ,
\end{equation}
where

\begin{equation}
H_{2,I}^{^{\prime }}=\eta _k(e^{i\theta }\left| J,-J+k+1\right\rangle
\left\langle J,-J+k\right| +e^{-i\theta }\left| J,-J+k\right\rangle
\left\langle J,-J+k+1\right| ).
\end{equation}
The interaction induces the transition
\begin{equation}
\left| J,-J+k\right\rangle \rightarrow \cos (\eta _{k+1}t)e^{-i\alpha
_kt}\left| J,-J+k\right\rangle -ie^{i(\theta -\alpha _{k+1}t)}\sin (\eta
_{k+1}t)\left| J,-J+k+1\right\rangle .
\end{equation}

Suppose that we desire to generate the superposition state
\begin{equation}
\left| \psi _d\right\rangle =\sum_{k=0}^Kd_k\left| J,-J+k\right\rangle ,
\end{equation}
where $d_k$ is a complex number, i.e., $d_k=\left| d_k\right| $ $e^{i\varphi
_k}$. Without lose of generality, we here assume that $d_0$ is real, i.e., $%
\varphi _0=0$. Assume that the qubit system is initially in the state $%
\left| J,-J\right\rangle $, i.e., all the qubits are initially in the ground
state. We divide the time interval into K subintervals. The duration of the
kth subinterval is $t_k$. During the kth subinterval, the frequency of the
classical driving source is $\omega _k=\omega _0+2(J-k)\lambda $. The
corresponding phase is $\theta _k$. After the first subinterval the system
evolves to

\begin{equation}
\left| \psi _1\right\rangle =\cos (\eta _1t_1)\left| J,-J\right\rangle
-ie^{i(\theta _1-\alpha _1t_1)}\sin (\eta _1t_1)\left| J,-J+1\right\rangle .
\end{equation}
Adjust the duration $t_1$ to satisfy the following condition
\begin{equation}
\cos (\eta _1t_1)=d_0.
\end{equation}
Then we obtain
\begin{equation}
\left| \psi _1\right\rangle =d_0\left| J,-J\right\rangle -i\sqrt{1-\left|
d_0\right| ^2}e^{i(\theta _1-\alpha _1t_1)}\left| J,-J+1\right\rangle .
\end{equation}
After the second subinterval the system evolves to
\begin{equation}
\left| \psi _2\right\rangle =d_0\left| J,-J\right\rangle -i\sqrt{1-\left|
d_0\right| ^2}e^{i(\theta _1-\alpha _1t_1)}[\cos (\eta _2t_2)e^{-i\alpha
_1t_2}\left| J,-J+1\right\rangle -ie^{i(\theta _2-\alpha _2t_2)}\sin (\eta
_2t_2)\left| J,-J+2\right\rangle ]
\end{equation}
Setting
\begin{equation}
\sqrt{1-\left| d_0\right| ^2}\cos (\eta _2t_2)=\left| d_1\right| ,
\end{equation}
we have
\begin{equation}
\left| \psi _2\right\rangle =d_0\left| J,-J\right\rangle -i\left| d_1\right|
e^{i[\theta _1-\alpha _1(t_1+t_2)]}\left| J,-J+1\right\rangle -\sqrt{%
1-\left| d_0\right| ^2-\left| d_1\right| ^2}e^{i(\theta _1+\theta _2-\alpha
_1t_1-\alpha _2t_2)}\left| J,-J+2\right\rangle .
\end{equation}
After each subinterval the highest excitation number is increased by 1. The
length of the kth subinterval $t_k$ satisfy
\begin{equation}
\sqrt{1-\sum_{j=0}^{k-1}\left| d_j\right| ^2}\cos (\eta _kt_j)=d_k.
\end{equation}
This leads to the final state
\begin{equation}
\left| \psi _K\right\rangle =d_0\left| J,-J\right\rangle
+\sum_{k=1}^Kc_k\left| J,-J+k\right\rangle ,
\end{equation}
where
\begin{eqnarray}
c_k &=&\left| d_k\right| e^{i\phi _k}, \\
\phi _k &=&\sum_{j=1}^k\theta _j-\alpha
_k\sum_{j=1}^{K-k+1}t_j-\sum_{j=1}^{k-1}\alpha _jt_j-k\pi /2.  \nonumber
\end{eqnarray}
Choose the phase of the driving source appropriately so that
\begin{equation}
\theta _k=\varphi _k-\sum_{j=1}^{k-1}\theta _j+\alpha
_k\sum_{j=1}^{K-k+1}t_j+\sum_{j=1}^{k-1}\alpha _jt_j+k\pi /2.
\end{equation}
Then we obtain $\phi _k=\varphi _k$ and $\left| \psi _K\right\rangle $ is
just the desired state $\left| \psi _d\right\rangle $. In the following we
propose an implementation of the idea with a cavity QED system, but it is
not restricted in cavity QED.

\section{PHYSICAL IMPLEMENTAION IN CAVITY QED}

We consider N two-level atoms interacting with a quantized cavity field and
driven by a weak classical field. The Hamiltonian is
\begin{equation}
H=H_f+H_{a-q}+H_{a-c},
\end{equation}
where
\begin{equation}
H_f=\omega _0S_z+\omega _ca^{+}a,
\end{equation}
\begin{equation}
H_{a-q}=g(a^{+}S^{-}+aS^{+}),
\end{equation}
\begin{equation}
H_{a-c}=\varepsilon [e^{-i(\omega t-\theta )}S^{+}+e^{-i(\omega t-\theta
)}S^{-}],
\end{equation}
$a^{+}$ and $a$ are the creation and annihilation operators for the cavity
field, $\omega _0,$ $\omega _c,$ and $\omega $ are the frequencies for the
atomic transition, cavity mode, and classical field, g is coupling constant
between the atoms and the cavity field, and $\varepsilon $ and $\theta $ are
the Rabi frequency and phase of the classical field. In the case $\delta
_c=\omega _0-\omega _c\gg g\sqrt{\stackrel{-}{n}+1}$, with $\stackrel{-}{n}$
being the mean photon number of the cavity field, there is no energy
exchange between the atomic system and the cavity. Then the Hamiltonian $%
H_{a-q}$ can be replaced by the effective Hamiltonian [10]

\begin{equation}
H_e=\lambda _c[\sum_{j=1}^N(\left| e_j\right\rangle \left\langle e_j\right|
-\left| g_j\right\rangle \left\langle g_j\right| )a^{+}a+S^{+}S^{-}],
\end{equation}
where $\lambda _c=g^2/\delta _c$. The first and second terms describe the
photon-number dependent Stark shift, and the last term describes the dipole
coupling between the atoms induced by the cavity mode. When the cavity mode
is initially in the vacuum state $\left| 0\right\rangle $ it will remain in
the vacuum state throughout the procedure. Then the Hamiltonian $H_f$
reduces to $H_0$ of Eq. (2) and $H_e$ reduces to $H_1$ of Eq. (3). The
parameters $\omega ,$ $\varepsilon $, and $\theta $ are controllable by the
classical field.

We now discuss the experimental implementation of the proposed scheme. In
recent cavity QED experiments with long-living Rydberg atomic levels coupled
to a cavity mode, the coupling constant is $g=2\pi \times 25kHz$ [11,12].
The atomic radiative time and photon damping time are about $T_r=3\times
10^{-2}s$ and $T_c=1\times 10^{-3}s$, respectively. Set $N=3$, $\delta =10g$%
, and $\varepsilon =g/100$. Suppose that the desired state is
\begin{equation}
\left| \psi _d\right\rangle =\frac 1{\sqrt{2}}(\left| 3/2,-3/2\right\rangle
+\left| 3/2,-1/2\right\rangle ).
\end{equation}
Then the required atom-cavity-field interaction time is $t=\pi /(4\eta
_1)=\pi /(4\sqrt{3}\varepsilon )\simeq 0.29\times 10^{-3}s.$ In this case
the decay time for the superposition state $\left| \psi _d\right\rangle $ is
$T_d=T_r/3=10^{-2}s.$ As the cavity mode is only virtually excited the
effective decoherence rate due to the cavity decay is $\kappa =g^2/T_c\delta
^2=$ $10Hz$. The infidelity induced by the decoherence is on the order of $%
t/T_d+t\kappa =3.19\times 10^{-2}$.

We now consider the probability that the atomic system undergoes a
transition to the state $\left| 3/2,1/2\right\rangle $ via the off-resonant
coupling. The detuning between the classical field and the transition $%
\left| 3/2,-1/2\right\rangle \rightarrow \left| 3/2,1/2\right\rangle $ is $%
2\lambda =0.2g$. The probability that the atomic system undergoes this
transition is given by

\begin{equation}
P_{-1/2\rightarrow 1/2}\sim \frac 12\frac{\eta _{k+1}^2}{\eta
_{k+1}^2+\lambda ^2}\sin ^2(\sqrt{\eta _{k+1}^2+\lambda ^2}t)\simeq
0.38\times 10^{-2}.
\end{equation}
With all of the above mentioned nonideal situations being considered, the
total error is about $3.57\times 10^{-2}$.

\section{SUMMARY}

In conclusion, we have discussed a physical mechanism for arbitray control
of quantum states for multi-qubit in the symmetric Dicke subspace. As a
consequence of qubit-qubit coupling, the energy levels are not equidistant
in the symmetric Dicke subspace. If the driving field is tuned in resonance
with a specific transition in the symmetric Dicke subspace, it would be
off-resonant with other transitions. This allows one to selectively
manipulate a prechosen transition. Any symmetric entangled state can be
created by adjusting the driving field. The Hamiltonian can be realized in
physical systems with qubit-qubit coupling available. We propose an
implementation of the idea in cavity QED. The entanglement of symmetric
Dicke states is robust against the loss of particles, as demonstrated in a
recent experiment with photons [13]. The idea opens new perspectives for
research of entanglement properties of general symmetric multi-qubit states.
The idea can also be used for preparation of entangled Dicke states for two
atomic systems. Suppose that the first atomic system is first prepared in a
superposition of Dicke states through the above mentioned procedure. Then
this atomic system is entangled with a light field via the exchange of
excitations induced by atom-field interaction. The two atomic systems can be
prepared in an entangled Dicke state by transferring the excitations of the
light field to the second atomic system [14]. This provides a possibility
for tests of quantum nonlocality with two entangled atomic systems. Apart
from fundamental tests of quantum theory, entangled Dicke states are useful
for quantum communication [15].

This work was supported by the National Natural Science Foundation of China
under Grant No. 10674025 and the Doctoral Foundation of the Ministry of
Education of China under Grant No. 20070386002.

\end{document}